# Prediction of SINR using BER and EVM for Massive MIMO Applications

Tim W. C. Brown[1], David A. Humphreys[2], Martin Hudlicka[3], Tian H. Loh[2]

[1] Institute for Communication Systems, Guildford, Surrey, GU2 7XH, UK, tim@brown76.name
[2] National Physical Laboratory, Hampton Road, Teddington, Middlesex TW11 0LW, UK, david.humphreys@npl.co.uk, tian.loh@npl.co.uk
[3] Czech Metrology Institute, Okruzni 31, 63800 Brno, Czech Republic, mhudlicka@cmi.cz

*Abstract*—Future communication systems employing massive multiple input multiple output will not have the ability to use channel state information at the mobile user terminals. Instead, it will be necessary for such devices to evaluate the downlink signal to interference and noise ratio (SINR) with interference both from the base station serving other users within the same cell and other base stations from adjacent cells. The SINR will act as an indicator of how well the pre-coders have been applied at the base station. The results presented in this paper from a 32 x 3 massive MIMO channel sounder measurement campaign at 2.4 GHz show how the received bit error rate and error vector magnitudes can be used to obtain a prediction of both the average and dynamically changing SINR.

## I. Introduction

In the emerging developments within 5G standardisation, Massive Multiple-Input Multiple-Output (mMIMO) is a high priority case requiring estimation of signal to interference and noise ratio (SINR). It is not practical to obtain the channel-state information (CSI) at the mobile due to the large array of antenna elements employed at the base station not allowing MIMO feedback. Instead, pilot signals are sent up to the base station from all mobiles to gain channel state estimation such that they could apply a zero-forcing pre-coder to reach multiple users, assumed to have single or multiple antennas [1]. The pilot signals are potentially subject to contamination from neighbouring cells as illustrated in Fig. 1, which can cause inaccuracy of the channel estimations resulting in pre-coder error.

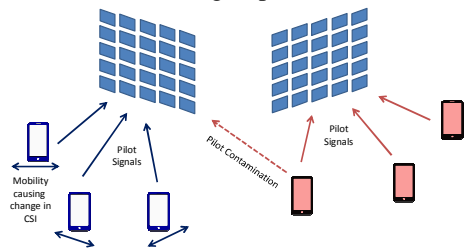

Fig. 1 – Illustration of pilot signals sent to a massive MIMO base station

Once the pre-coder is applied as shown in Fig. 2, the resulting antenna radiation patterns will cause interference to other users as well as communicate to the intended user. Therefore, this can be considered as inter user interference. At the same time, pre-coders from neighbouring cells will also cause an inter-cell interference, which is envisaged to be significantly less. The pre-coder will become outdated as the mobiles move such that they exceed their coherence distance in the pre-coded channel, therefore a second cause of inter-cell interference is mobility. It is necessary for each mobile in real time to obtain periodic SINR estimations and report back a channel quality indicator (CQI) [2] as illustrated in Fig. 2. This will subsequently inform the base station about whether the pre-coder is sufficiently updated in order to produce a sufficient SINR to each user based on the reported CQI values.

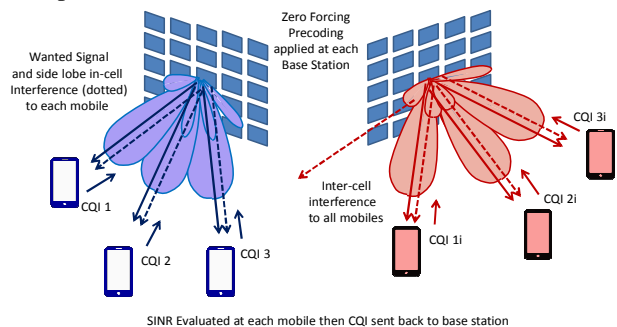

Fig. 2 – Application of the zero forcing pre-coder with inter-user and inter-cell interferences

The relation between bit error ratio (BER) and signal to noise ratio (SNR) have been well established for different propagation environments [3], while at the same time capacity has been seen as a way to predict SINR [4], though it is also very dependent on the channel state. For high SNRs above 10 dB, BER drops substantially in a time invariant or slowly varying propagation channel, which is difficult to measure reliably. This will subsequently make such SINR levels difficult to predict with such a metric. On the other hand, when the dominant noise is at the receiver, error vector magnitude (EVM) is well known to have an inverse square root proportionality to SNR [5], which could apply to SINR, but the effect of different noise characteristics have to be understood. The weakness, however, is that it would require additional signal processing on top of the digital receiver to implement, which does take place at baseband. This paper details some wideband channel sounder measurement results carried out to show the ability to track SINR both in frequency and space using EVM as the means to predict.

## II. MODEL DERIVATION FOR PREDICTING SINR

It is first necessary to show by simulation of synthetic data how the BER and EVM can be used to form a model whereby the SINR can be predicted from the output BER or EVM. It is necessary to perform simulations with synthetically generated data using real wave forms to create interference that cannot be modelled simply by Gaussian noise. An initial simulation is carried out comprising a 20 MHz, 1200 carrier 64 QAM orthogonal division frequency multiplex (OFDM) signal transmitted over a flat-fading channel and a co-channel 64 QAM OFDM interferer. The interference and noise were normalized to return the SINR value. The results for the BER and root mean square (RMS) EVM are evaluated over 1200 carriers and 20 transmitted frames on each carrier. The results of the BER and the RMS EVM are shown in Fig. 3 (a) and (b) respectively.

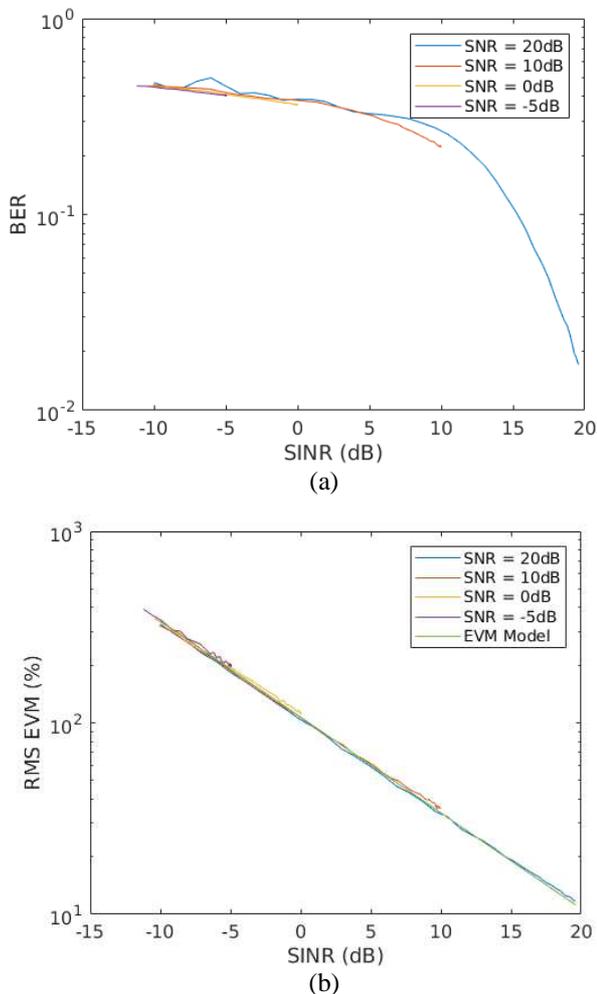

Fig. 3 – Initial simulated result showing the relationship of (a) BER and (b) EVM to SINR.

The results in Fig. 3 are evaluated for fixed SNR values between -5 dB and 20 dB, but the SINR reaches a maximum equivalent to the SNR in all cases, where there is zero interference. It should be noted that these results are unaffected by the number of carriers and so the number chosen is arbitrary. As with SNR, the BER does drop rapidly at SINR values of 10 dB and above, while it is nearly flat for such levels below 10 dB and this makes it hard to create a simple linear relationship to the SINR. EVM on the other hand, shows a clear log linear relationship and a model function has been inserted into Fig. 3 (b) which will be studied in this work. Therefore, a relationship can be made between the linear SINR and the EVM as follows:

$$\text{EVM}(\%) = \frac{A}{\sqrt{\text{SINR}}} \quad (1)$$

Consequently, the predicted SINR, $\text{SINR}_P$ in dB can be predicted from the EVM as follows:

$$\text{SINR}_P \text{ (dB)} = 20\log_{10}\left(\frac{A}{\text{EVM}(\%)}\right) \quad (2)$$

The value of the gradient, determined by $A$ is highly dependent on the QAM order. Some repeated simulations had been undertaken to show this comparison where an example is given in Fig. 4 of how the EVM to SINR relationship changes with different QAM order. The simplest 4 QAM is substantially different where SINR cannot be modelled as a single interferer and this will cause a consistent shift in the quadrature phase resulting in a near constant EVM. For higher orders of QAM, the relationship can be modelled and the value of $A$ for up to three interferers and different QAM orders is shown in Table I.

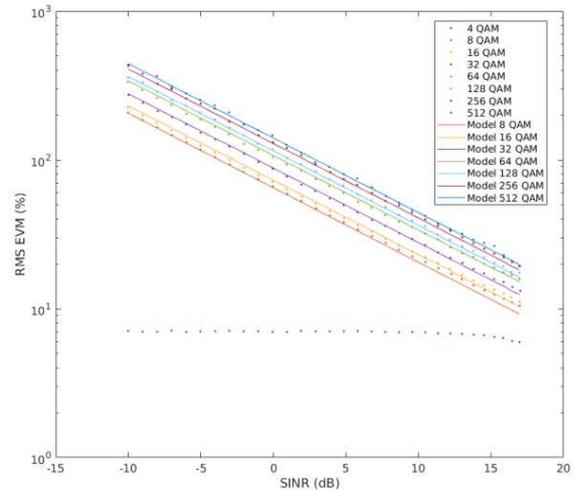

Fig. 4 – Simulated results showing the impact of QAM order on SINR prediction for OFDM using a single interferer

Table I - Comparison of the gradient value for the log linear relationship between EVM and SINR for differing QAM order and up to three interferers (Ints).

| QAM Order | $A$ (1 Int) | $A$ (2 Ints) | $A$ (3 Ints) |
|---|---|---|---|
| 8 | 65 | 77 | 77 |
| 16 | 73 | 78 | 78 |
| 32 | 88 | 90 | 92 |
| 64 | 107 | 107 | 107 |
| 128 | 115 | 115 | 115 |
| 256 | 129 | 129 | 129 |
| 512 | 140 | 140 | 140 |

It is clear from the results in Table I that regardless of the number of interferers, a QAM order of 64 or above yields the same model when using 20 frames (as in this simulation) or above. This is expected as the higher order of QAM becomes more sensitive to interference that changes in the number of interferers superimposing on the wanted signal and this will have a similar random effect on the vector error. Additionally, the increase in QAM order is critical to ensuring a de-correlation between the wanted and interfering signal, which reduces to less than 2% with a chosen QAM order of 64 in this work.

Finally, it is necessary to justify the 20 iterations or frames used to calculate the RMS EVM. Fig. 5 (a) illustrates a plot of EVM vs SINR, not plotted logarithmically for clarity, with fixed numbers of iterations. It can be seen that there is some deviation from the model for just 2 iterations and this reduces as the number of iterations increases.

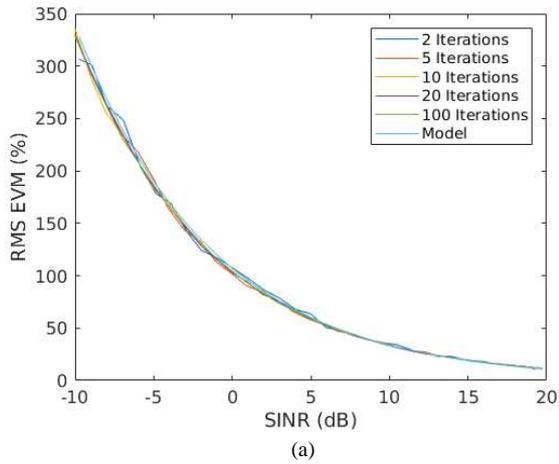

(a)

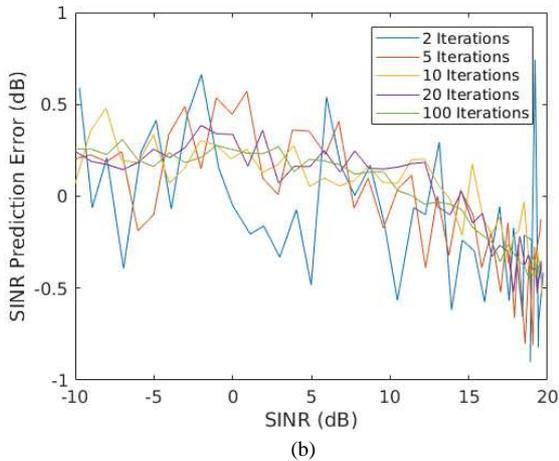

(b)

Fig. 5 – Comparison of (a) the EVM to SINR and (b) the SINR prediction error for different values of SINR

It can be concluded from Fig. 5 (a) that an increased number of iterations will produce a more repeatable EVM for a given SINR. The required repeatability, and hence minimum required number of iterations, is useful to evaluate by defining the prediction error, which is the difference between $SINR_P$ and the actual simulated SINR with signalling. As the EVM is more repeatable with a higher number of iterations, the gap will close and hence prediction error reduces. Fig. 5(b) shows the prediction error vs simulated SINR and it is clear that with more than 10 iterations, the prediction error is within ±0.5dB. Hence 20 iterations are chosen based upon this criteria to lead to reliable reporting of CQI.

## III. SUMMARY OF MEASUREMENT SETUP

A 32x6 MIMO wideband channel sounder setup was used to undertake the measurements over a 200 MHz bandwidth (of which 120 MHz was used for evaluation) at 2.4 GHz. Two 32x3 massive MIMO measurements were effectively taken simultaneously as it is required to have greater than ten times the number of elements at the transmitter than the number of single antenna receivers to meet the criteria [6]. The measurement was carried out in an outdoor obstructed line of sight environment with significant vegetation to enable some channel fading but varying separation of beam space between the three receivers as illustrated in Fig. 6 as a plan view with a photograph of the environment in Fig. 7. The transmit array consisted of 32 dual polar patch elements, while the receiver antennas Rx1 to Rx 6 were commercial omnidirectional monopole antennas with a small ground plane attached to them. For the first set of 32x3 data, Rx1 to Rx3 were used, where their close proximity, stationarity and fixed position shown in Fig. 6 would create substantial interference thus limiting SINR. For the second set of 32x3 data, there is more beam-space separation between Rx4 and Rx5, while Rx6 is moving as indicated in Fig 6 to yield variable SINR for tracking.

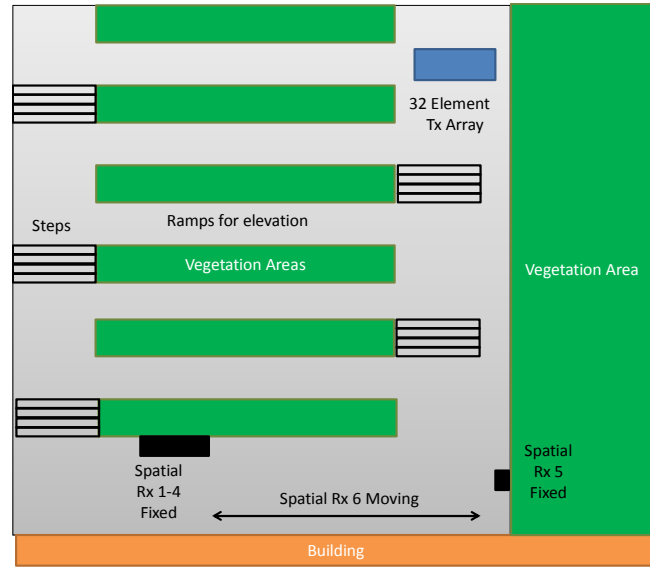

Fig. 6 – Plan view of the outdoor massive MIMO measurement area

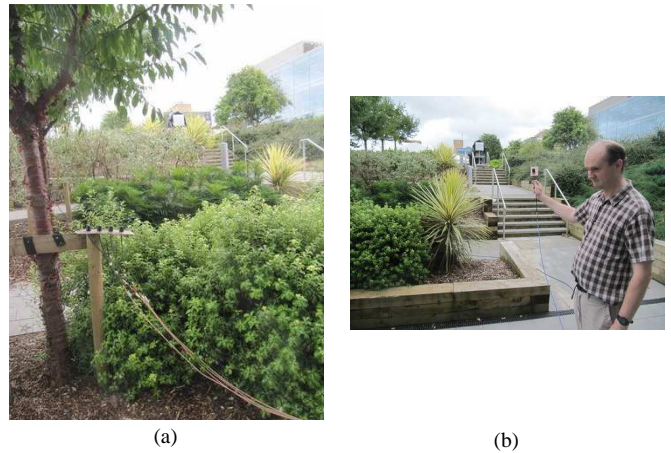

(a)          (b)

Fig. 7 – Photographs of the measurement environment showing the 32 element transmit array and (a) Rx1 to Rx4 at a fixed position and (b) Rx 6 moving

Measurements were carried out lasting approximately 20 seconds, where the sample to sample time period was 36.56 ms corresponding to a 27.35 Hz sampling rate. This was selected as it was found to be more than three times the maximum Doppler frequency at a walking speed of 4 km/h, which was 8.89 Hz. Hence the coherence time of the channel could be sufficiently determined, while any possibility of aliasing regarding the second order channel statistics was avoided. The channel impulse responses were measured and stored for post processing.

## IV. MEASUREMENT POST PROCESSING RESULTS

In post processing, using the measured channel data, a zero forcing pre-coder [3] was applied to the base station elements such that each of the three receivers would have a corresponding pre-coded channel, but interference would be caused by the pre-coded channels of the other two receivers. In this setup, the SINR with signalling $SINR_S$, is defined as the ratio of the power from the wanted pre-coded channel transmitting a wanted OFDM signal, $E[var(s_w)]_f$, to the interfering pre-coded channels transmitting independent OFDM signals, $E[var(s_{i1})]_f$ and $E[var(s_{i2})]_f$ plus Gaussian additive noise variance, $\sigma^2$ as follows:

$$SINR_S = \frac{E[var(s_w)]_f}{E[var(s_{i1})]_f + E[var(s_{i2})]_f + \sigma^2} \quad (3)$$

Note here that the term $E[var(s_x)]_f$ is defined such that for each OFDM carrier, the variance is evaluated. Subsequently the mean or expected value of the variances over all carriers is computed. To compare $SINR_S$ with $SINR_P$ in this measurement setup, OFDM signals with 2 MHz bandwidth and 120 carriers were chosen and hence there were 60 OFDM signals over the whole 120 MHz bandwidth used to evaluate the SINR in both time and frequency.

Fig. 8 (a) shows the mesh plot of $SINR_S$ with the corresponding $SINR_P$ in Fig. 8 (b) for receiver Rx3. Note the peak SINR is below 10 dB as Rx1 to Rx3 are closely located and stationary, thus they have substantial interference making it hard for the pre-coder to create beam-space between them. The difference between $SINR_P$ and $SINR_S$ constitutes the prediction error in this scenario and this corresponds to the mesh plot in Fig. 8 (c), which should ideally have a flat response of 0 dB. This analysis using real measured propagation channel data is necessary to determine whether the frequency selectivity (or non frequency flat) channel has any substantial impact on the SINR prediction. The total RMS EVM will be different to that of a flat fading channel as individual carriers in the OFDM signal transmitted within a deep fade within the channel will result in substantially high EVM. The prediction error can reach values beyond ±0.5 dB unlike the simulations using synthetic data on a frequency flat channel using the same number of iterations.

Fig. 9 carries out the same analysis as in Fig. 8 but for Rx6, so that the impact of a moving receiver can be evaluated. Therefore, in this case the SINR has more time variant and due to wider antenna separation can reach higher SINR values above 10 dB, while also varying between -10 dB and 20 dB as shown in Fig. 9 (a) and (b). This therefore gives more indication of the ability to use OFDM frames to carry out dynamic tracking of a time variant SINR. Prediction error was unaffected due to this time variant change as shown in Fig. 9 (c).

It is worthy to note that while wanted and interfering signals are de-correlated, which is crucial to generating repeatable results; it is also important to consider the repeatability of the variance in the signalling such that the actual value of $SINR_S$ is repeatable when transmitted over the same propagation channel. This was evaluated for each channel state whereby a ±2.15% variance over 500 repeated blocks was evaluated in each block containing 20 OFDM frames. For low SINRs, where additive Gaussian noise is a negligible contribution, such variance would result in an error in $SINR_S$ of just ±0.19 dB.

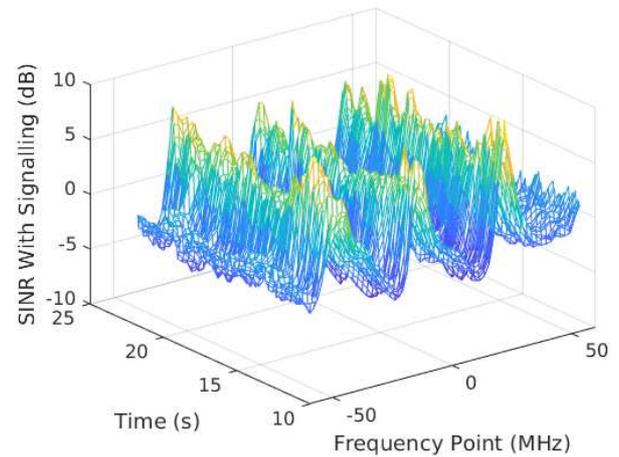
(a) $SINR_s$ with signalling

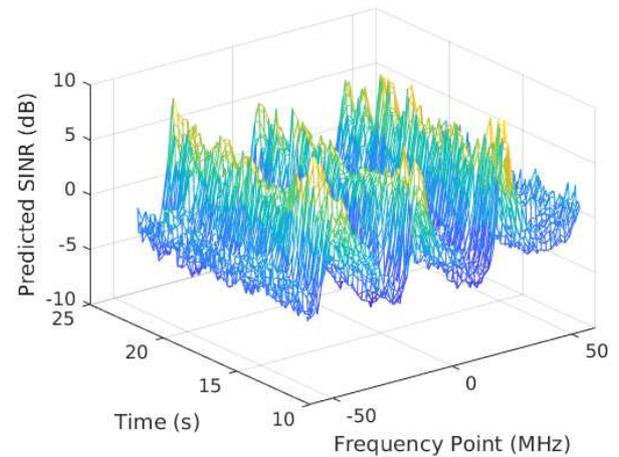
(b) Predicted $SINR_P$

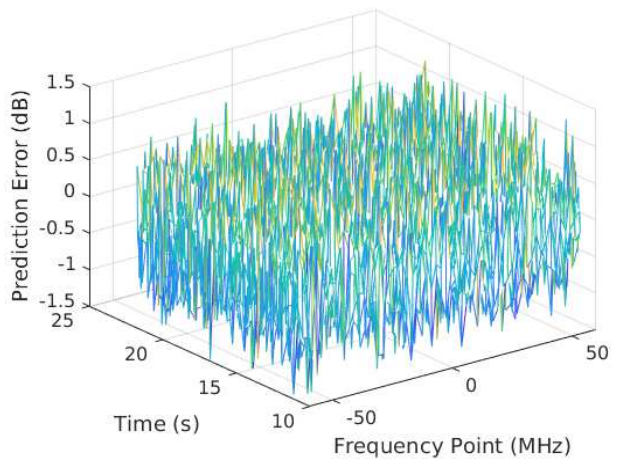
(c) Prediction error

Fig. 8 – A comparison of the measured (a) and predicted (b) SINR for a stationary receiver (Rx3) shows good agreement (c) across the 120MHz bandwidth.

Finally, the impact on prediction error due to the bandwidth of the OFDM signal for Rx3 and Rx 6 is shown in Fig. 10. First, it should be noted that both Rx3 and Rx6 cases present a low mean and standard deviation on the prediction error with the OFDM bandwidth of 2MHz used for the two example cases in Fig. 8 and Fig. 9.

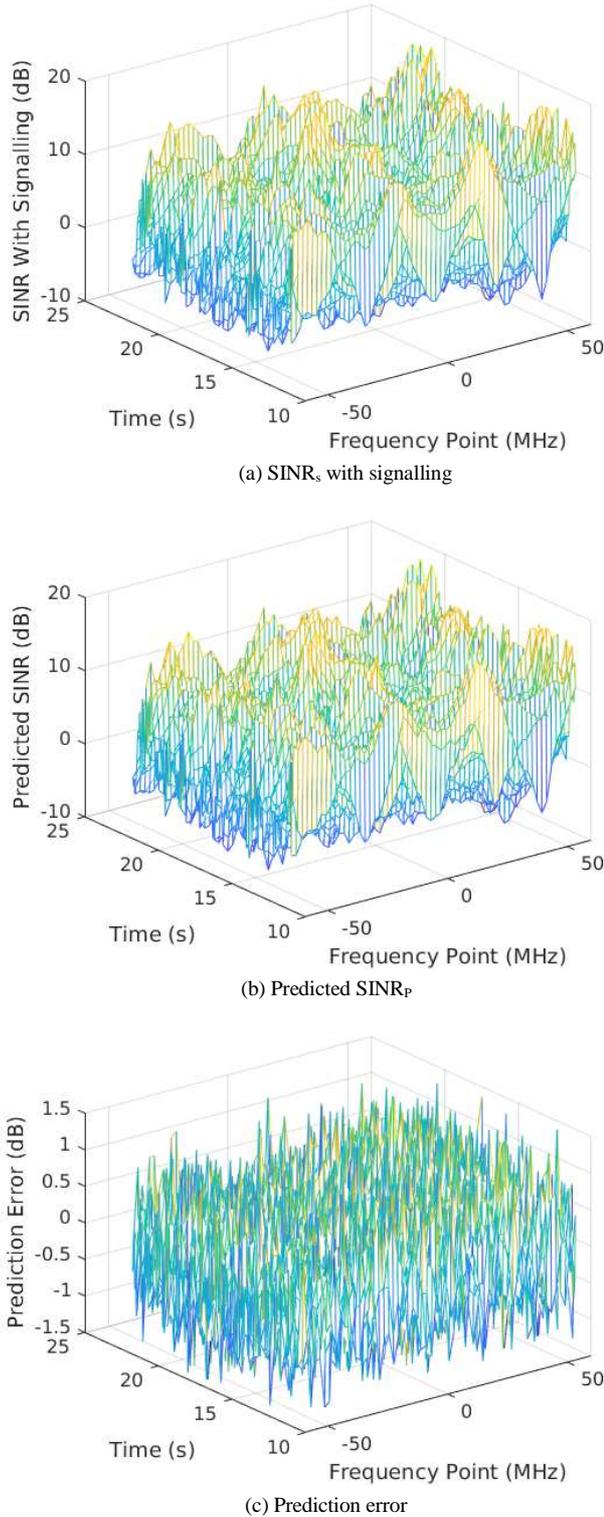

(a) SINR$_s$ with signalling

(b) Predicted SINR$_P$

(c) Prediction error

Fig. 9 – A comparison of the (a) measured and (b) predicted SINR for a moving receiver (Rx6) shows good agreement across the 120 MHz bandwidth

It is useful to evaluate in Fig. 10 how this mean and standard deviation changes with OFDM bandwidth (and proportionately the number of carriers), since wider bandwidths with more frequency selectivity will invite a greater prediction error. Clearly bandwidths between 2MHz and 5MHz allow a low standard deviation on the prediction error, while increasing the bandwidth substantially worsens this, reaching a higher error for Rx6 than Rx3 due to the wider frequency selectivity of that channel going beyond bandwidths where the actual SINR is coherent. Another interesting observation is that a very narrowband OFDM of 1MHz has a higher standard deviation than 2MHz. This could be explained by the fact that a small number of carriers cause greater uncertainty in the RMS EVM.

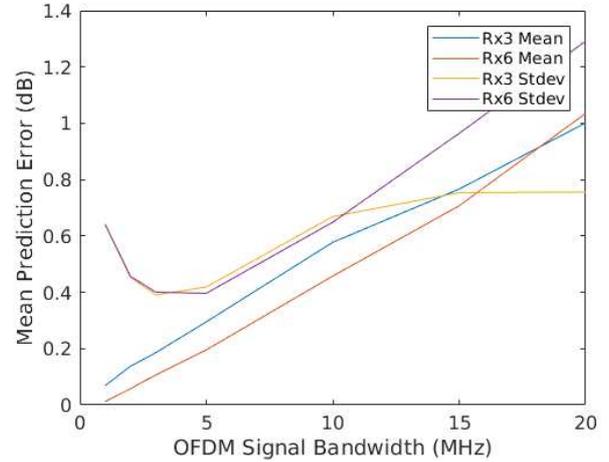

Fig. 10 – Mean prediction error evaluated vs OFDM channel bandwidth

## V. CONCLUSION

Results of a massive MIMO measurement campaign were evaluated to indicate the capability of EVM to predict SINR where the sources of interference and their magnitude are unknown. The prediction of SINR can be used for a more accurate and reliable CQI, dependent on low prediction error on the SINR with signalling. Such prediction error in a real channel was found not to exceed ±2 dB, though this is dependent on the bandwidth of the OFDM pilots used and the number of frames to predict the SINR based on a corresponding RMS EVM. The bandwidth of such pilots should be within the coherence of the SINR in the channel.

## ACKNOWLEDGEMENT

The results in this paper come from the project EMPIR 14IND10 MET5G (Metrology for 5G Communications). This project has received funding from the EMPIR programme co-financed by the Participating States and from the European Union's Horizon 2020 research and innovation programme.